\documentclass[%
 aip,
 amsmath,amssymb,
 reprint,%
 nofootinbib
]{revtex4-1}

\usepackage{graphicx}
\usepackage{dcolumn}
\usepackage{bm}
\usepackage{enumerate}

\usepackage{setspace,color,url}
\allowdisplaybreaks[1]

\usepackage[dvipsnames]{xcolor}

\newcommand\revb[1]{\textcolor{black}{{#1}}}

\usepackage{graphicx}
\usepackage{soul}
\usepackage{mathrsfs}

\newcommand{\dC}[1]{\frac{\partial}{\partial C}\left(#1\right)}
\newcommand{\dr}[1]{\frac{\partial}{\partial r}\left(#1\right)}
\newcommand{\drr}[1]{\frac{\partial^2 #1}{\partial r^2}}
\newcommand{\Th}[1]{\mathrm{th}\left( #1\right)}

\newcommand{\dt}[1]{\frac{\partial #1}{\partial t}}
\newcommand{\dx}[1]{\frac{\partial #1}{\partial x}}

\newcommand{\rev}[1]{\textcolor{black}{#1}}

\begin{document}

\preprint{Submitted to The Journal of Chemical Physics}


\title{Active sieving across driven nanopores for tunable selectivity}

\author{Sophie Marbach}
\author{Lyd\'eric Bocquet}
\affiliation{Laboratoire de Physique Statistique, UMR CNRS 8550, Ecole Normale Sup\'erieure, PSL Research University, 24 rue Lhomond, 75005 Paris, France}
\email{lyderic.bocquet@lps.ens.fr}
\date{\today}
\begin{abstract}
Molecular separation traditionnally relies on sieving processes across passive nanoporous membranes. Here we explore theoretically the concept of non-equilibrium active sieving. We investigate a simple model for an active \rev{noisy} nanopore, where gating - in terms of size or charge - is externally driven at a tunable frequency. Our analytical and numerical results unveil a rich sieving diagram in terms of the forced gating frequency.
Unexpectedly, 
the separation ability is strongly increased as compared to its passive \rev{(zero frequency)} counterpart. It points also to the possibility of tuning dynamically the osmotic pressure. Active separation outperforms passive sieving and represents a promising avenue for advanced filtration.

\end{abstract} 
\maketitle

\section{Introduction}

Filtering specific molecules is a challenge faced for numerous vital needs: from biomedical applications like dialysis to the intensive production of clean water.~\cite{Elimelech2011,Bocquet2014,Elimelech2016} Most modern processes for filtration are based on passive sieving principles: a membrane with specific pore properties allows to separate the permeating components from the retentate. The domain has been boosted over the last two decades by the possibilities offered by nanoscale materials, such as graphene or advanced membranes.~\cite{Noskov2004,Karnik2011,Karnik2014,Bakajin2006,Geim2014,Siria2013,Fulinski2002,Picallo2013,Feng2016} Selectivity requires small and properly decorated pores at the scale of the targeted molecules, and this inevitably impedes the flux and transport, making separation processes costly in terms of energy. These traditional sieving membranes are also passive, therefore unable to adapt to external changes, like varying salt or contaminant concentrations in the liquid to filtrate. Furthermore while Nature is able to distinguish quasi similar ions, \textit{e.g.} like sodium and potassium,~\cite{Greger1983} no artificial counterpart has been designed up to now to reach such a fine selectivity. 

In this context we explore the possibility of {\it  active sieving}, harnessing non-equilibrium dynamics to
separate particles across nanopores. A Maxwell demon is the (utopian) prototypical system able to perform separation on the basis of transfer of information.~\cite{Szilard1929} However
designing active pores that can distinguish between nanometer-scale molecules presents the obvious challenge of measuring \textit{in situ} the proper information, {\it i.e.} fabricating feedback nanocontrollers.~\cite{Stone2014,Koski2014,Bechhoefer2014}
Now one may consider a simpler situation of an active nanopore that can change its transmission properties with time thanks to an external energy input.   
This corresponds accordingly to a non-equilibrium situation,  baring some analogy with active matter, which allows to bypass to some extent
the equilibrium constraints for better separation.

\begin{figure}[h]
\center
\includegraphics[width=7.5cm]{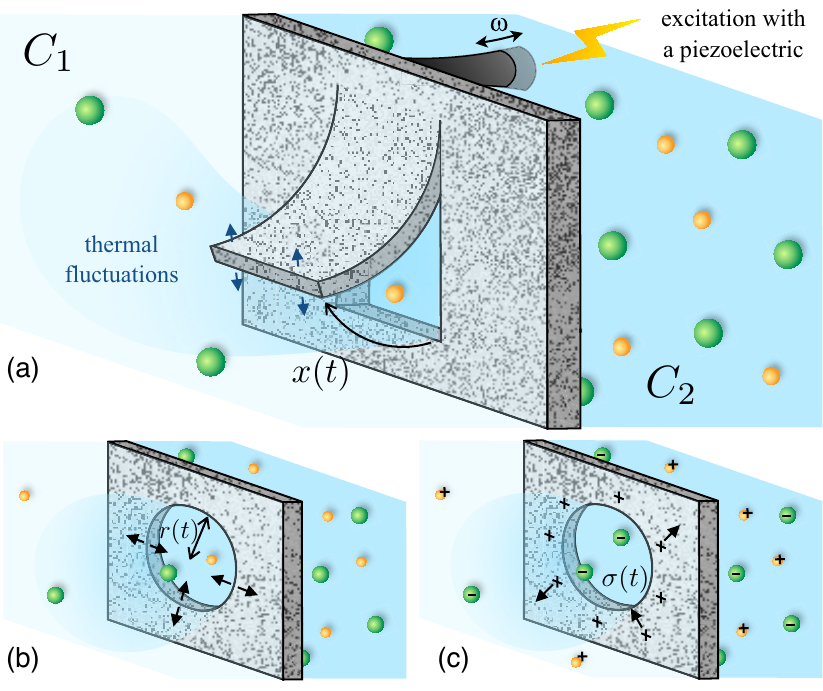}
\caption{a - A driven nanodoor in a membrane , submitted to opening by thermal fluctuations and by external application of a periodic excitation. b - Circular nanopore with fluctuating size and c - with fluctuating charge.}
\label{fig1} 
\end{figure}

Here we explore a simple situation, where an external mechanical or electrical action modifies the pore properties, 
thus creating some blind -- "crazy" -- Maxwell demon.
Typical geometries of driven nanopores under consideration are sketched in Fig.~\ref{fig1}: a driven nano-gate; a pore with a fluctuating size; or a pore whose surface charge may be externally gated. Such geometries are of special interest in the present study since they are 
 amenable to further experimental investigations.
 %
To model separation across these systems, we build on the pioneering work R. Zwanzig in Refs.~\citenum{Zwanzig1990, Zwanzig1992}, who considered the translocation rate of molecules through fluctuating pores. 
We consider as a supplementary ingredient that the opening of the nanopore is forced externally at a given frequency $\omega$. 
%

\vspace{2mm}

\section{An active pore model} 

\subsection{Effusion of solute through an active pore}

We consider the effusion of a solute (with concentration $C$ in a reservoir) across a nanopore.
Pore gating which controls the translocation state across the nanopore is characterized by an internal parameter $x$: for example the radius of the pore, the door opening, 
or the surface charge, see Fig.~\ref{fig1}. In line with Zwanzig's model of Refs.~\citenum{Zwanzig1990, Zwanzig1992}, we assume that the solute concentration $C$
relaxes according to the following leakage equation:
\begin{equation}
\frac{dC}{dt} = - K(x) C ,
\label{eqC}
\end{equation}
where $K(x)$ is the $x$-dependent leakage constant. It is proportional to the mobility of the solute.
It also depends on the characteristics of the gating. 
For steric gating, $x$ is merely geometrical: for the circular pore in Fig.~\ref{fig1}b, $x$ is the radius $r$ of the pore and $K(r) = k' r^2$, while for the nanodoor $x$ is the aperture of the door and $K(x) = k \vert x\vert$. For electric gating -- when the pore is charged, see Fig.~\ref{fig1}-c, $x$ is proportional to the surface charge of the pore $\Sigma$ and for small nanopores one may model $K(x) = k"\sqrt{1+x^2}$ (see Appendix A for details). 
The constant $k$ (resp. $k'$ and $k''$) defines the mobility of the solute.

\revb{We are interested in the separation of small sized particles (say, ions, colloids, polymers, ...)
effusing through nanometric sized pores.} 
Accordingly the internal parameter $x$ is further assumed to evolve dynamically due to (i) thermal noise \revb{-- expected at the nanoscale --}, and (ii) some external forcing which drives an oscillation.
To simplify the discussion, we assume that the nanopore is excited such that its {\it average} opening $x$ oscillates at a frequency $\omega$,
and we further model the effects of thermal noise 
by a simple Langevin equation for the excess internal parameter $\delta x = x - \langle x\rangle$:
\begin{equation}
\begin{split}
\label{eqNoise}
 \langle x \rangle &= x_0(t) = x_0\sin(\omega t)  \\
\frac{d \, \delta x}{dt} &= - \lambda\delta x + F(t)
\end{split}
\end{equation}
where $F(t)$ is a gaussian white noise. The second moment of $\delta x$ is $\left< \delta x^2 \right> = \theta$, and the fluctuation-dissipation theorem at equilibrium imposes $\left< F(t) F(t') \right> = 2\theta\lambda\delta(t-t')$.  The goal is now to obtain more information on the evolution of the solute concentration averaged over the noise:  $\langle C(t) \rangle$.

\subsection{From a rate process to the Schmoluchowski equation}

We turn to the equivalent Fokker-Planck -- or Smoluchowski -- equation for $C$.
This derivation is inspired by Ref.~\citenum{Zwanzig1990}. We denote $f(C,x,t)$ the probability distribution that the variables $C$ and $x$ have specified values at time $t$. This function satisfies a Liouville equation:
\begin{equation}
\label{eqLiouville}
\dt{f} = - \dC{\frac{dC}{dt}f} - \dx{\frac{dx}{dt}f}
\end{equation}
or, putting in the velocities explicitly, 
\begin{equation}
\begin{split}
\label{eqLiouville2}
\dt{f} =  & - \dC{-K(x)Cf} \\
& - \dx{\left(-\lambda (x-x_0) f + \frac{dx_0}{dt} f + F(t)f\right)}.
\end{split}
\end{equation}
Now we would like to average this stochastic Liouville equation to have the average of $f$ over the noise: $g(C,r,t) = \left< f(C,r,t) \right>_{\mathrm{noise}}$. We may rewrite the Liouville equation in terms of an operator $L$ such that Eq.~(\ref{eqLiouville2}) is:
\begin{equation}
\dt{f} = - Lf - \dx{F(t)f}.
\end{equation}
It integrates into:
\begin{equation}
f(C,r,t) = e^{-tL}f(C,x,0)-\int_0^t ds e^{-(t-s)L}\dx{F(s)f}
\end{equation}
that we use to rewrite the differential equation as:
\begin{equation}
\dt{f} = - Lf - \dx{F(t)\left(e^{-tL}f(C,r,0)-\int_0^t ds e^{-(t-s)L}\dx{F(s)f}\right)}.
\end{equation}
Now we can safely average over the noise, \revb{using the gaussian properties of $F(t)$, namely $\langle F(t) \rangle = 0$ and $ \langle F(t) F(t') \rangle  = 2\theta \lambda \delta(t-t')$,} which gives:
\begin{equation}
\dt{g} = - Lg + \dx{\lambda\theta \dx{g}}
\end{equation}

Now we look for the average value of $C$ at time $t$ and key feature $x$: $\bar{C}(x,t) = \int dC C g(C,x,t)$. This yields the following differential equation (Schmoluchowski equation):
\begin{equation}
\label{Schmo2}
\dt{\bar{C}} = - K(x)\bar{C} + \revb{ \dx{} \lambda\theta \dx{\bar{C}}} + \dx{ \left(\lambda (x - x_0) - \frac{dx_0}{dt}\right) \bar{C}}
\end{equation}

\subsection{Permeance of the active pore}

\begin{figure}[h]
\center
\includegraphics[width=8.0cm]{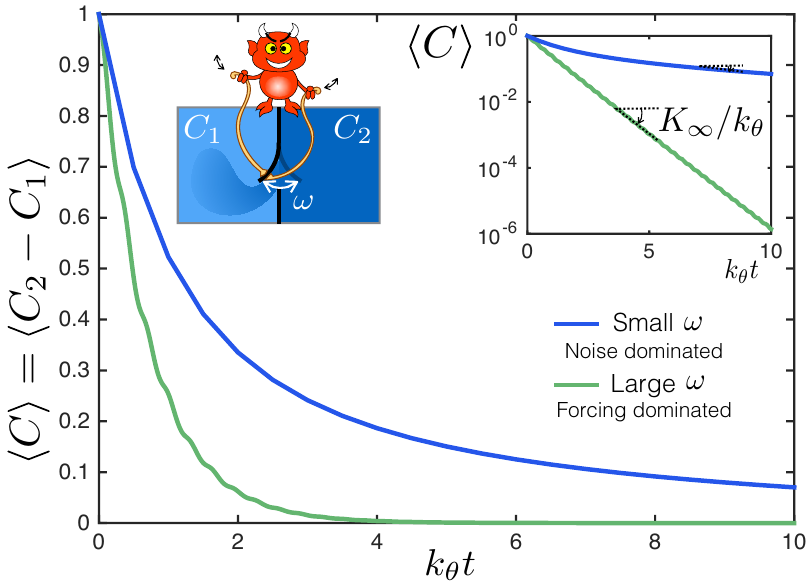}
\caption{Relaxation of the difference in concentration between the two sides of a nanodoor, averaged over noise. The data is a simulation result with $\lambda/k_{\theta} = 10^{-3}$, $\omega/k_{\theta} = 10^{-3}$, and $\omega/k_{\theta} = 1$. (\textit{Inset, Left}) Illustration of a demon oscillating the nanodoor. (\textit{Inset, Right}) Log scale of the previous graph and example of the extraction of the long time relaxation constant $K_{\infty}$, the \textit{permeance} of the system.}
\label{fig0} 
\end{figure}

The time-dependent concentration  $\langle C(t) \rangle$ is accordingly defined as $\int \overline{C}(x,t) dx  = \langle C(t) \rangle $. \revb{For simplicity in the following we consider that the noise damping parameter $\lambda$ does not depend on $x$.} 
The Smoluchowski equation Eq.(\ref{Schmo2}) can be solved analytically for some specific forms of $K(x)$ (in particular for $K(x) \propto x^2$). Alternatively
we solve Eq.(\ref{Schmo2}) numerically, to deduce the time dependent averaged concentration $\langle C(t) \rangle $.
We show in Fig.~\ref{fig0} an example for the averaged concentration $\left< C(t) \right>$, here in the case of a nanodoor where $K(x)=k \vert x\vert$
(Fig.~\ref{fig1}a).

As a generic feature, 
one may show that $\left< C(t) \right>$ is exponentially decaying at long times, 
\begin{equation}
\label{permeance}
\left< C(t) \right> \underset{t\rightarrow \infty}{\sim} \exp(-K_{\infty} t).
\end{equation}
and this allows to define the \textit{permeance} $K_{\infty}$ of the system. 
Fast translocation of the solute corresponds to a large $K_{\infty}$.
The permeance $K_{\infty}$ depends on the thermal damping $\lambda$, noise amplitude $\theta$, but also on the external forcing (frequency $\omega$ and amplitude $x_0$). In the following, our goal is to identify general rules on how the permeance depends on these antagonistic effects. 
Units of lengths are given by $\sqrt{\theta}$, while time is given in terms of a renormalized parameter $k_{\theta}$ with units of an inverse  time: for the nanodoor, $k_\theta= k\sqrt{\theta}$, while for the nanopore $k_\theta= k'\, {\theta}$.


\section{Transport through the active pore} 
\begin{figure}[b]
\center
\includegraphics[width=9cm]{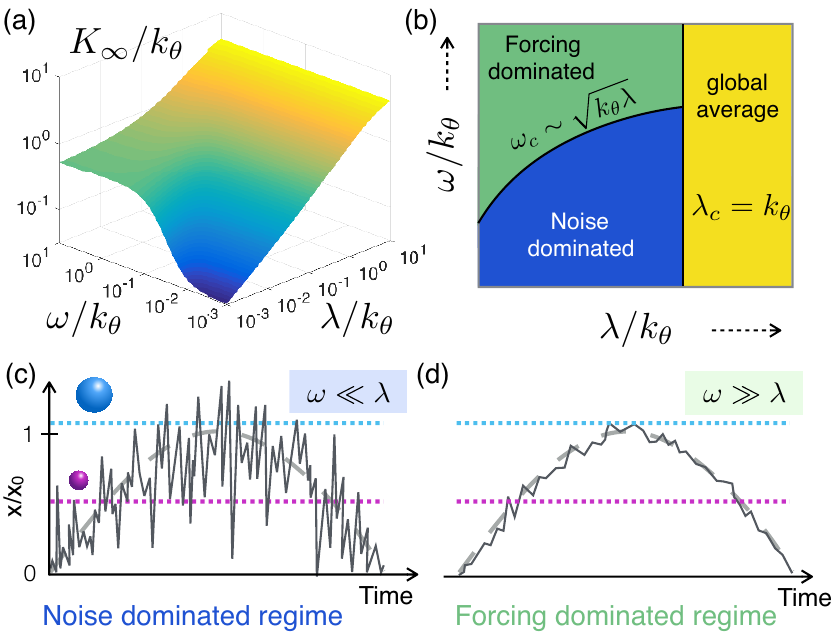}
\caption{a - Analytical solution of the permeance $K_{\infty}$ as a function of the forcing frequency $\omega$ and the thermal damping $\lambda$ for the forced nanopore system ($r_0 = 2\sqrt{\theta}$ and $\epsilon = 0.5$). b~-~Universal phase diagram of the permeance $K_{\infty}$ with $\omega$ and $\lambda$. c - Schematic of the door opening $x/x_0$ (solid black line) in the noise dominated regime, where $\omega \ll \lambda$ and d - in the forcing dominated regime, where $\omega \gg \lambda$.}
\label{fig2} 
\end{figure}

Let us first focus on oscillating circular pore, in which case 
the leakage law writes  $K(r)=k' r^2$,
with $r$ the pore radius, see Fig.~\ref{fig1}b. We study fluctuations around the averaged forced radius $\langle r(t) \rangle = r_0(1+ \epsilon \cos(\omega t))$ with a given amplitude $r_0\epsilon$. In this case, the Smoluchowski equation can be solved analytically (see Appendix B) and the 
expression for the permeance writes:
\begin{equation}
\begin{split}
K_{\infty}(\omega) &= \lambda/2 \left(\left(1+\frac{4k'\theta}{\lambda}\right)^{1/2}-1 \right) +  k'r_0^2\left(1+\frac{4k'\theta}{\lambda}\right)^{-1} \\
&+ k'\frac{r_0^2\epsilon^2}{2}\left( \frac{\left(1+ \frac{4k'\theta}{\lambda}\right)^{-1}}{1 + (\frac{\omega}{\omega_c(\lambda)})^2} + \frac{(\frac{\omega}{\omega_c(\lambda)})^2}{1 + (\frac{\omega}{\omega_c(\lambda)})^2} \right)
\end{split}
\label{Kinf}
\end{equation}
with
$\omega_c(\lambda)  = \sqrt{4k'\theta \lambda + \lambda^2}$ a cut-off frequency.
The first term of Eq.~(\ref{Kinf}) corresponds to the solution for the non-forced case studied by Zwanzig in Ref.~\citenum{Zwanzig1992}. In his derivation, only fluctuations of a bottleneck opening are considered \revb{(with a hard reflecting barrier at $r = 0$ so that only positive radii are considered)}. With $r_0 = 0$ and $\omega = 0$ one recovers exactly the exponential factor of Eq.~(8) of Ref.~\citenum{Zwanzig1992}. The last term corresponds to the supplementary leakage induced by the forced oscillations: it is the combination of a low-pass and a high-pass filter. 
The general behavior of $K_{\infty}(\omega)$ is plotted in Fig.~\ref{fig2}-a. It exhibits complex features that are summarized in the diagram of Fig.~\ref{fig2}-b.
Although it is presented here only for the nanopore, the diagram is generic to all the systems investigated and represented in Fig.\ref{fig1}.

Let us discuss the various regimes at play. It is first interesting to explore the 
limiting behaviors at low damping. \rev{This regime is actually relevant for ionic or liquid separation systems,~\cite{Noskov2004,Secchi2016} see for instance the experimental study of biological channels in Ref.~\citenum{beece1980} which is consistent with the low damping limit relaxation with $\lambda \propto 1/\eta$, with $\eta$ the fluid viscosity}. For low and high frequencies, we can calculate from Eq.~(\ref{Kinf}) (at highest order):
\begin{equation}
\begin{split}
& K_{\infty}(\omega) \underset{\underset{\lambda  \ll k_\theta}{\omega \ll \omega_c}}{\sim} \sqrt{ k' \lambda \theta} \\
& K_{\infty}(\omega)  \underset{\underset{\lambda  \ll k_\theta}{\omega \gg \omega_c}}{\sim}k' r_0^2\epsilon^2/2
\end{split}
\label{LimitBehaviors}
\end{equation}
These results call for a generic physical interpretation. 
At high frequency, the forced oscillations become too quick for the thermal damping 
to rub them out and $K_{\infty}$ reduces simply to its noise average: $K_{\infty} \simeq (\omega/2\pi) \int K[x_0 (t)] dt$. 
This is the \textit{forcing dominated regime}, see Fig.~\ref{fig2}d. 
The behavior at low forcing frequencies $\omega$, where noise dominates (see Fig.~\ref{fig2}c), is more subtle. According to Eq.~(\ref{eqNoise}), the gating
variable will mainly diffuse with a diffusion coefficient ${\cal D}_x=\theta\,\lambda$. Over a time $\tau$, the gating variable thus takes a typical value $\bar{x}\sim \sqrt{{\cal D}_x \tau}$. Now the passage time is itself fixed by $K_{\infty}^{-1}$, so that one gets a self-consistent estimate for $K_{\infty}$,
as 
\begin{equation}
K_{\infty}  \underset{\underset{\lambda  \ll k_\theta}{\omega \ll \omega_c}}{\simeq} K\left[\bar{x}\sim\sqrt{\frac{\lambda \theta}{K_{\infty}}}\right]
\label{SC}
\end{equation}
For the circular nanopore, where $x$ is the radius $r$ and $K(r)=k' r^2$, one deduces accordingly $K_{\infty}\approx \sqrt{k'\,\lambda \theta}$ as obtained in Eq.(\ref{LimitBehaviors}). 
This interpretation for $K_{\infty}$ in Eq.~({\ref{SC}) can be generalized to the other types of gates. For the nanodoor, for which $K(x)=k \vert x\vert$,
Eq.~({\ref{SC})  predicts $K_{\infty} \approx (k \lambda\theta)^{1/3}$, as can indeed be verified numerically (see Appendix C).

The transition between the low and high frequency regimes results from the competition between the forced oscillations and the noise. 
In Eq.~(\ref{eqNoise}) the thermal fluctuations $f(t)$ compete with the forced oscillations $dx_0(t)/dt \sim \omega x_0(t)$ and the crossover between the two regimes occurs accordingly when $f(t) \sim dx_0(t)/dt$. Using the fluctuation-dissipation theorem, and taking a typical time-scale $\tau \sim k_\theta^{-1}$ this yields 
$\omega_c(\lambda) \sim \sqrt{k_{\theta}\lambda}$ for the critical frequency. This estimation matches perfectly the scaling obtained numerically for all  systems of Fig.~\ref{fig1} and also with the full analytical expression Eq.~(\ref{Kinf}) for the circular pore of Fig.~\ref{fig1}b.

%

\vspace{2mm}

\section{Dynamical selectivity}

\subsection{Dynamical gating on mobility}

The different scalings in Eq.~(\ref{LimitBehaviors}) suggest further that the passage rate $K_{\infty}$ exhibits a strongly contrasted dependence on the particle mobility (via $k'$) in the low and high frequency regimes. 
Accordingly, at finite frequency, solutes with different mobilities will be separated by the active gate in a very different way as compared to the static (passive) nanopore.

This is highlighted in Fig.~\ref{fig3}a-c, where we show the permeance of the nanopore to particles of different permeabilities, corresponding to particles with different $k'$, (here $k_1/k_2 = 100  $ for illustration). The selectivity of the pore, defined in terms of the ratio of the permeances of the two particles, is plotted in Fig.~\ref{fig3}c. What is striking in this plot is that
the selectivity is a strongly dependent function of the frequency (and furthermore non monotonous), so that the relative translocation rate of the two species can be finely tuned by the forcing frequency. This stems from the fact that the critical frequency $\omega_c$ for each particle is dependent on the particle mobility (via $k'$). Thus, a slower-diffusing particle will reach the forcing dominated regime at smaller frequencies. When the slower (blue) particle has just transitioned to the forcing dominated regime, the faster (purple) particle is still in the noise dominated regime, and the selectivity is reduced. This points to various non trivial avenues for 'on demand' sieving. 
\begin{figure}[h!]
\center
\includegraphics[width=8.7cm]{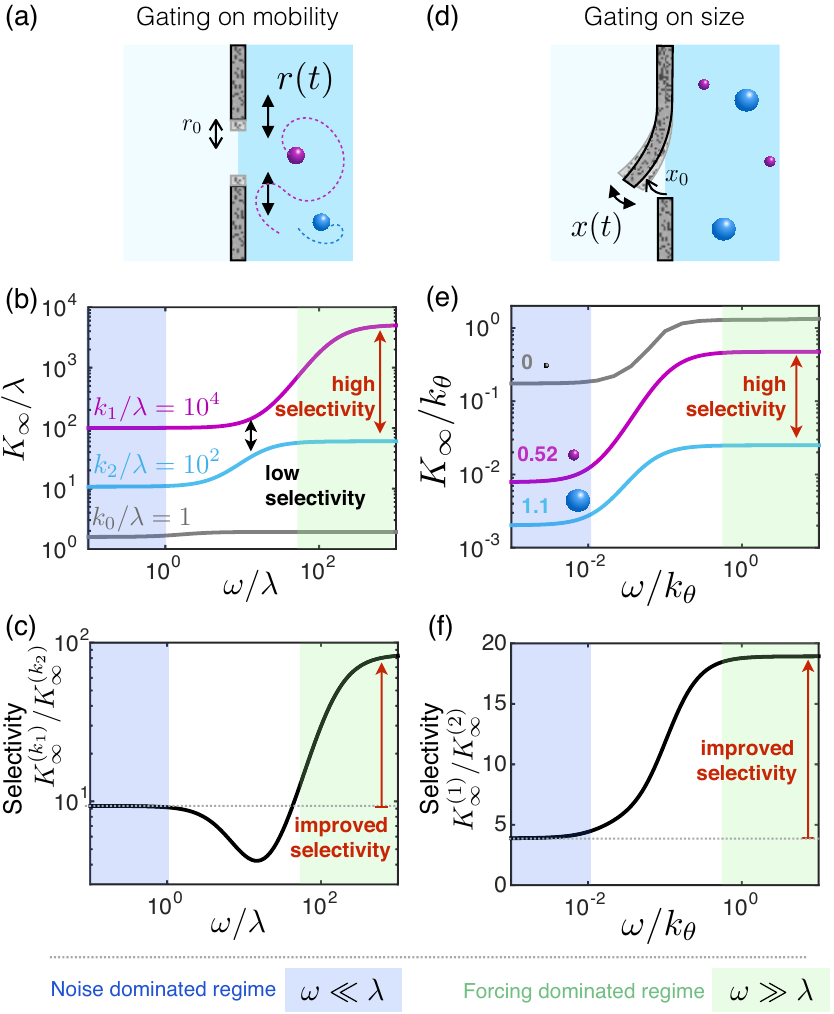}
\caption{ a - Schematic of gating through the fluctuating pore relying on mobility differences between particles. b - Permeance $K_{\infty}/\lambda$ through the nanopore of two particles with mobility $k_{1}/\lambda = 10^4$ and $k_{2}/\lambda = 10^2$ as a function of the forcing frequency $\omega/\lambda$, for small $\lambda$. c - Selectivity of the nanopore to those particles, defined as the ratio of the permeances. d - Schematic of gating through the fluctuating door relying on size differences between particles. e - Permeance $K_{\infty}/k_{\theta}$ through the nanodoor of two particles of different size (the smallest, purple is $0.52 x_0$ in radius and the largest, blue, $1.1 x_0$), as a function of the forcing frequency $\omega/k_\theta$, for small $\lambda$. f - Selectivity of the nanodoor to those particles, defined as the ratio of their permeances.}
\label{fig3}
\end{figure}

\revb{We emphasize that these results are not dependent on the choice of relative mobility, and here $k_1/k_2 = 100$ is chosen for readibility. In a more realistic case of ionic separation, for instance separating sodium and potassium, we would have $k^{(K^+)}/k^{(Na^+)} = 1.47$.~\cite{Handbook} As a consequence, for low frequencies the selectivity $K_{\infty}^{(K^+)}/K_{\infty}^{(Na^+)} \sim \sqrt{k^{(K^+)}/k^{(Na^+)}} \sim 1.21$ and at high frequencies the selectivity increases: $K_{\infty}^{(K^+)}/K_{\infty}^{(Na^+)} \sim k^{(K^+)}/k^{(Na^+)} \sim 1.47$. Note that this does not depend on the value of the noise damping parameter $\lambda$: as long as noise is signficant in the system, one will always find the critical frequencies from one regime to another.}


\vspace{2mm}

\subsection{Dynamical gating on size}

This behavior is generic to all gatings described in Fig.~\ref{fig1}. To highlight this generic feature we conclude by considering the dynamical selectivity of the
nanodoor, represented in Fig.~\ref{fig1}-a, with a slightly
modified gating process taking explicitly in consideration 
the effect of the finite size of the particle, see Fig.~\ref{fig3}-d. We use a similar gating function as in Ref.~\citenum{Eizenberg1995},
 so that particles cannot pass if the opening $x$ of the pore is smaller than their size $x_p$. We modify accordingly 
 the leakage law of Eq.~(\ref{eqC}) to $K(x-x_p) \mathcal{H}(x-x_p)$ where $\mathcal{H}$ is the Heaviside function. 
For this leakage law, the Smoluchowski equation  cannot be solved analytically and we turn to numerical solutions, see Appendix C.
In Fig.~\ref{fig3}-e  we compare the measured permeance for three particle sizes: an infinitely small particle (size 0 in grey), a small particle (in purple) and a large particle (in blue). As above, we deduce the corresponding selectivity factor for the two particles with different size as the ratio of their permance.
As obtained above for the other gating processes, we find a selectivity that is dependent on the frequency, here a strongly increasing function of the frequency.

\section{ Discussions and Conclusions }

\revb{\subsection{General Conclusion}}

These results show that the selectivity of nanoporous membranes can finely be tuned by an externally forced gating.
Depending on the forcing frequency, dynamical gating allows 
to better discriminate particles with different size or mobility. 
As a rule, in the limit of low damping common in liquid or ionic filtration, an active pore is thus capable of filtering more precisely smaller particles than a standard passive filter with fixed pore size (at zero frequency). Also, in the high frequency regime, an active pore sieves particles in terms of their mobility, which is interesting to separate particles with similar size or charge (as would be needed for the separation of ions, for instance for distinguishing sodium and potassium that have similar size and charge but different mobility)~\cite{Handbook}. 
Although simple, our model provides a rich diagram, highlighting noise dominated or forcing dominated regimes, with specific selectivity rules.
 These selectivity properties may be tuned by adjusting the frequency of the excitation, and rely on the strong interplay between noise and external excitation.

{Numerous extensions can be obviously  proposed for the model, which we now plan to  explore exhaustively.}
\revb{The model could be easily exploited to explore the consequences of several extensions. If the noise damping parameter $\lambda$ now depends on $x$, one expects the critical frequency $\omega_c$ and the limiting regimes to be modified in a non-trivial way. Furthermore, since the equations are not linear, when a non-monochromatic excitation is triggered, mode coupling will occur, and may result into a broader variety of behaviors.}
\revb{Another underlying questions in the prospect of possible applications of this research -- in particular within the field of desalination and filtration -- is that of the energy consumption of such a device. Obviously the active or dynamical part of the sieving requires some energy input, however that energy input depends on the specific means of excitation and a detailed energy balance is required to predict the energy efficiency of such dynamical sieving process, a question which we leave for future work. 
}

 {However the present results already suggest a number of developements for experimental implementations of active pores.}
 Nanoporous materials with piezoelectric or piezomechanical response, {\it e.g.} metal organic frameworks,~\cite{fu2008,ortiz2014} are promising candidates in this goal. 
Furthermore, a nanodoor like in Fig.~1a can be designed by nanofabrication techniques, {\it e.g.} carving membranes at the sub-micron scale using a focused ion beam. 
Forcing at a tunable frequency, as well a supplementary white noise, can be provided by piezoelectric systems, allowing to explore the various domains in the dynamical sieving diagram. 
These possibilities are a few examples for experimental realizations. They constitute natural routes for a proof of concept of the ideas presented here. 

\revb{\subsection{Towards an on-demand osmotic pressure}}

We conclude with a final comment on osmotic pressure. 
\revb{As highlighted by Kedem and Katchalsky in the context of membrane transport, there is an intimate symmetry link between permeance and osmosis.~\cite{Kedem} }
A non-vanishing (resp. vanishing) osmotic pressure is expressed for a semi- (resp. fully) permeable membrane.
This link is highlighted by the generic expression for the osmotic pressure~\cite{Kedem}
\begin{equation}
\Delta \Pi = \sigma \times k_BT \langle C \rangle 
\end{equation}
introducing the rejection coefficient $\sigma$, whose value is equal to 1 (resp. 0) for a semi- (resp. fully) permeable membrane{; for a finite  permeance $K_{\infty}$ of the membrane, one then expects $1-\sigma \propto K_{\infty}$.~\cite{Kedem} 
Going to dynamical sieving, the pore opening occurs intermittently with the frequency $\omega$ supplemented by thermal noise, so that an intermittent osmotic pressure builds up. 
 Let us explicit this link using the extended model with steric gating, with leakage law $K(x-x_p) \mathcal{H}(x-x_p)$ leading to an effective permeance $K_{\infty}(\omega\vert x_p)$. The corresponding solute flux $J_s =K_{\infty}\, \mathcal{V}\,   \langle C \rangle$ (where $\mathcal{V}$ is the volume of the reservoir) can be identified to its definition
$J_s=\bar{\mathcal{A}} \frac{D}{e}\kappa\langle C\rangle$ where $ \bar{\mathcal{A}} $ 
is the average opening area of the pore, $D$ the diffusion coefficient of the solute and $e$ the thickness of the membrane;  the permeability coefficient $\kappa$ is accordingly related to $\sigma$ as $\kappa \propto 1 - \sigma$.~\cite{Kedem}
Gathering definitions, one thus obtains the dynamical rejection coefficient in terms of selectivity:
\begin{equation}
\sigma(\omega) = 1 - \frac{K_{\infty}(\omega\vert x_p)}{K_{\infty}(\omega\vert x_p=0)}
\end{equation}
where the permeance of a particle with vanishing size $x_p=0$ is used as normalization. 
This leads to a frequency dependent osmotic pressure,
$\Delta \Pi (\omega)= \sigma(\omega) \times k_BT \langle C \rangle $.
\revb{Note that this expression for the osmotic pressure is pertinent on time-scales longer than the time-dependent forcing.}

Our previous results for $K_{\infty}(\omega\vert x_p)$ show that $\Delta \Pi (\omega)$ is a strongly dependent function of $\omega$ via active sieving. \revb{This frequency dependence of the osmotic pressure is illustrated in Fig.~\ref{fig4} for various solutes.
Tuning the frequency of the forcing therefore allows to modify 'on demand' the osmotic pressure across the active membrane. 
This opens new avenues in terms of separation for active and `on-demand' reverse osmosis.}
\begin{figure}[h!]
\center
\includegraphics[width=8.7cm]{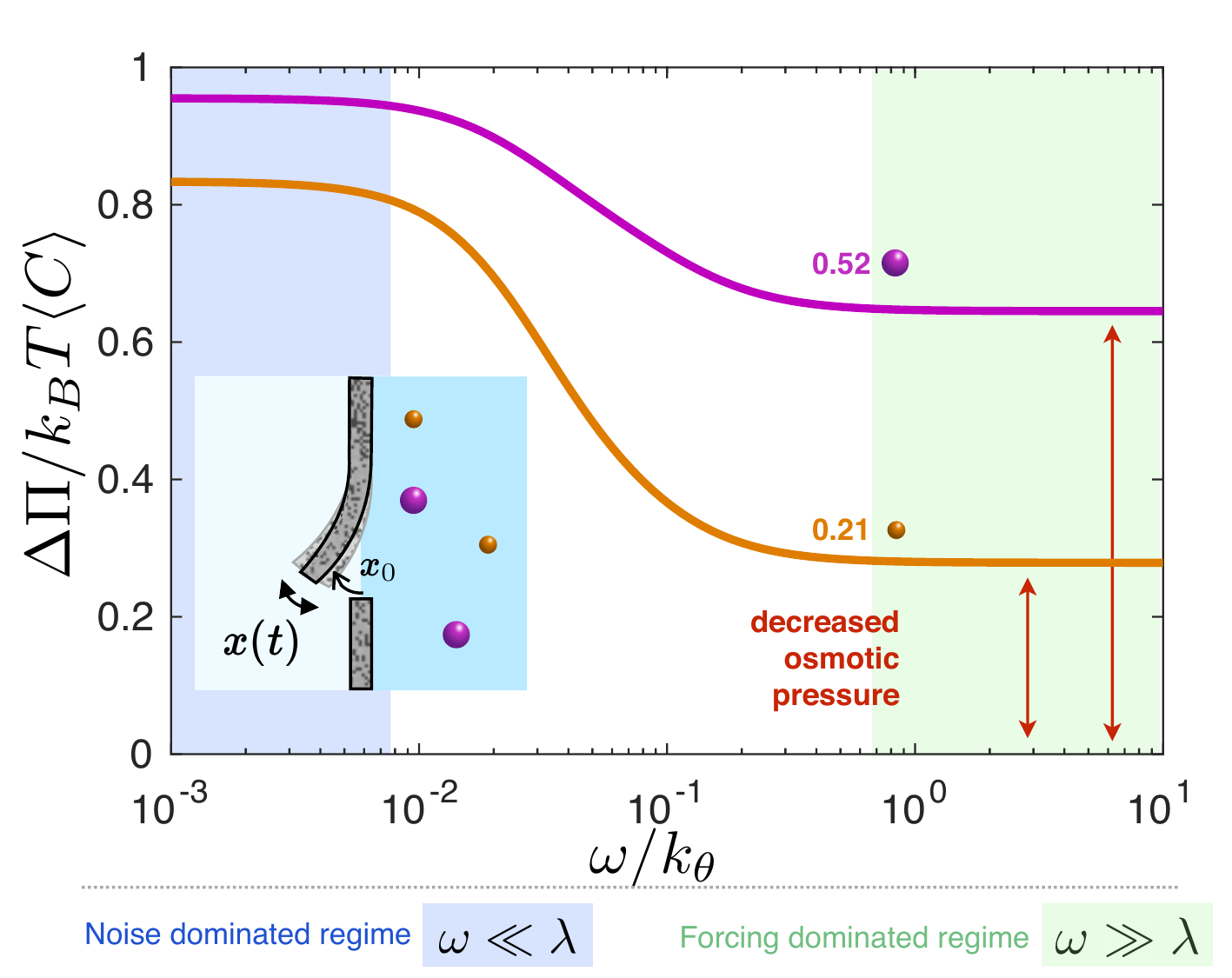}
\caption{\revb{
Frequency dependent osmotic pressure through the nanodoor for two particles of different size (the largest, purple, is $0.52 x_0$ in radius and the smallest, orange is $0.21x_0$) as a function of the forcing frequency $\omega/k_{\theta}$ for small $\lambda$. Inset: schematic of gating through the fluctuating door relying on size differences between particles.}}
\label{fig4}
\end{figure}

\vspace{5mm}



\section*{Acknowledgments}
The authors thank R. Netz, A. Siria and V. Kaiser for valuable discussions.
S.M. acknowledges funding from a J.-P. Aguilar grant of the CFM foundation. 


\section*{Appendix}

\subsection*{Appendix A : Examples of rate process laws}

\subsubsection*{A - 1 - Strictly diffusion limited processes through pores}

The initially concentrated reservoir has volume $\mathcal{V}$. The concentration at the scale of the pore is equilibrating over the typical thickness $e$ of the pore with $\mathcal{A}$ the pore's apparent section. 
We can write
\begin{equation*}
\frac{d C}{d t} = - \frac{D}{ \mathcal{V}e}  \mathcal{A} C
\end{equation*}

In the case of the nanodoor, $\mathcal{A} \sim x (h + w)$ where $x$ is the opening depth of the door (see main paper fig.1), $h$ is the height of the door, and $w$ it's width. As a consequence, we find:
\begin{equation}
\frac{dC}{dt} = - k |x| C , \hspace{2mm} \rm{with} \hspace{2mm} k =\frac{D(h+w)}{ \mathcal{V}e} 
\end{equation}
for the case of the nanodoor. We find that $k$ is directly proportional to the mobility -- or the diffusion coefficient -- of the solute. 

In the case of the circular pore $\mathcal{A} = \pi r^2$ where $r$ is the radius of the circular pore. As a consequence we find:
\begin{equation}
\frac{dC}{dt} = - k' r^2 C , \hspace{2mm} \rm{with} \hspace{2mm} k' =\frac{D \pi}{ \mathcal{V}e} 
\end{equation}
for the case of the circular pore. This is the expression used in Ref.~\citenum{Zwanzig1992}.


\subsubsection*{A - 2 - Charge influenced rate process}

Transport in a narrow charged channel of apparent section $\mathcal{A}$ can be described within the one-dimensional Nernst-Planck model from Ref.~\citenum{bocquet2010}. For the linear response, the relation between generalized fluxes and potentials is expressed via a transport coefficient matrix:
\begin{align*}
\begin{pmatrix}
I \\
\Phi_t
\end{pmatrix}
&=
\frac{\mathcal{A}}{e}
\begin{pmatrix}
K & \mu_K \\
\mu_K & \mu_\mathrm{eff}
\end{pmatrix}
\cdot
\begin{pmatrix}
-\Delta V \\
-kT\Delta(\log C_s)
\end{pmatrix}\\
K &= 2\mu q^2\sqrt{C_s^2 + \left( \frac{\Sigma}{h} \right)^2}\\
\mu_\mathrm{eff} &= 2\mu \sqrt{C_s^2 + \left( \frac{\Sigma}{h} \right)^2}\\
\mu_K &= 2\mu q \frac{\Sigma}{h}
\;,
\end{align*}
where $\Sigma$ is the number of surface charge, $h$ is the channel height, $\mu=\beta D$ is the ionic mobility and $q$ the elementary charge.

We assume small initial concentration difference $\Delta C_s$ (around the value $C_s$) between two reservoirs of volume $\mathcal{V}$, therefore we approximate $\Delta(\log C_s) \simeq (\Delta C_s)/C_s$. We apply no voltage difference, therefore we focus on the equation:
\begin{align*}
\frac{\mathrm{d} \Delta C_s}{\mathrm{d} t} = \frac{2\Phi_t}{\mathcal{V}} &= -\frac{2\mathcal{A}}{e\mathcal{V}}\mu_\mathrm{eff}kT\Delta(\log C_s)\\
 &\simeq -\frac{4 D\mathcal{A}}{e\mathcal{V}}  \sqrt{1 + \left( \frac{\Sigma}{h C_s} \right)^2} \Delta C_s
\end{align*}
and finally:
\begin{equation}
\frac{\mathrm{d} C}{\mathrm{d} t} = -k'' \sqrt{1 + \sigma^2} C
\end{equation}
where we relabelled the variables ($\Delta C_s \rightarrow C$, $\Sigma/h C_s \rightarrow \sigma$) in the last equation and introduced the characteristic rate $k'' = 4 D\mathcal{A} / e\mathcal{V}$. Note that $\sigma$ is the Dukhin number for the channel.


\subsection*{Appendix B : Exact solution of the rate process for circular nanopores}
\label{sectionAnalytics}

We consider the case where an external force excites the radius of the pore at the frequency $\omega$ around a non-zero mean value, so that:
\begin{equation}
\label{eqNoise1}
\left< r(t) \right>_{noise} = r_0 (1 + \varepsilon\sin(\omega t))
\end{equation}
The Schmoluchowski Eq.~(\ref{Schmo2}) becomes:
\begin{equation}
\begin{split}
\label{SchmoZ1}
\dt{\bar{C}} = &- kr^2\bar{C} + \lambda\theta\drr{\bar{C}} \\
&+ \dr{\left[\lambda(r-r_0(1+\varepsilon\sin(\omega t))) - r_0\varepsilon\omega\cos(\omega t)\right]\bar{C}}
\end{split}
\end{equation}

We assume that the probability distribution $\bar{C}$ initially has its equilibrium value in the absence of leakage, which simply writes $\bar{C}(r,t=0) = \exp(-\frac{1}{2\theta}(r-r_0)^2)$.

We look for a solution writing: $\bar{C}(r,t) = \exp(a(t) + b(t)r - c(t)r^2)$
This yields the following system of equations:
\begin{subequations}
\begin{align}
&\dot{a}(t) = -2\lambda\theta c +  b^2\lambda\theta + \lambda - br_0(\varepsilon\omega\cos{\omega t} + \lambda(1+\varepsilon\sin{\omega t}))\label{Schmo2a} \\
&\dot{b}(t) = \lambda b - 4bc\theta\lambda + 2cr_0(\varepsilon\omega\cos{\omega t} + \lambda( 1 + \varepsilon\sin{\omega t}))  \label{Schmo2b}\\
&\dot{c}(t) = k - 4c^2\lambda\theta + 2\lambda c \label{Schmo2c}
\end{align}
\end{subequations} 

We begin with Eq.~(\ref{Schmo2c}) which has the general solution:
\begin{equation*}
c(t) = \frac{1}{4\theta}\left(1+ S \mathrm{tanh}\left(\lambda S (t-t_0)\right)\right),
\end{equation*}
where $t_0$ is a constant that can be computed thanks to initial conditions, and
$S = \left(1+ \frac{4k\theta}{\lambda}\right)^{1/2}$. The initial conditions prescribe:
\begin{equation*}
c(t=0) = \frac{1}{2\theta} = \frac{1}{4\theta}\left(1+ S \tanh{(-\lambda St_0)}\right), 
\end{equation*}
so that $\Th{ - \lambda S t_0} =  1/S$. Replacing this result in the expression for $c$ gives:
\begin{equation}
\label{SchmoSolvec}
c(t) = \frac{1}{4\theta}\left(\frac{2+(S+1/S)\tanh{\lambda S t}}{1+1/S\tanh{\lambda S t}}\right).
\end{equation}

The differential Eq.~(\ref{Schmo2b}) is solved using the simple trick to write $b(t) = b_0(t)b_1(t)$ where $b_0(t)$ verifies the time differential equation involving the terms depending on $b$ only:
\begin{equation}
\dot{b}_0(t) = -\lambda b_0 \left (\frac{1+S\tanh{\lambda S t}}{1+1/S\tanh{\lambda S t}}\right)
\end{equation}
And so $b_0(t) = 1/\left(S\cosh{\lambda S t} + \sinh{\lambda S t}\right)$.
The equation on $b_1$ is then:
\begin{equation}
\dot{b}_1(t) =  2c(t)r_0(\varepsilon\omega\cos{\omega t} + \lambda(1+\varepsilon\sin{\omega t}))/b_0(t)
\end{equation}
and we find the integration constant such that $b(t=0) = r_0/\theta$. A lengthy but straightforward calculation leads to the solution of this equation as:
\begin{equation}
\begin{split}
b(t) &= {\frac{r_0}{2 S \left(S^2 \lambda ^2+\omega ^2\right) (S \text{cosh}[S t \lambda ]+\sinh[S t \lambda ])}} \\
& ... \times \bigg(\left(-1+S^2\right) \left(\omega ^2+S^2 \lambda  (\lambda -\varepsilon  \omega )\right)+ \text{cosh}[S t \lambda ] \\
& ... \times \big(\left(1+S^2\right) \left(S^2 \lambda ^2+\omega ^2\right) + S^2 \left(S^2 -1 \right) \varepsilon  \lambda  \omega 
\text{cos}[\omega t ] \\
&+S^2 \epsilon  \left(\left(1+S^2\right) \lambda ^2+2 \omega ^2\right) \text{sin}[ \omega t]\big) \\
& + S \big(2 \left(S^2 \lambda ^2+\omega ^2\right)  +\left(-1+S^2\right) \epsilon  \lambda  \omega  \text{cos}[ \omega t] \\
&+\epsilon  \left(\omega
^2+S^2 \left(2 \lambda ^2+\omega ^2\right)\right) \text{sin}[\omega t]\big) \text{sinh}[S t \lambda ]\bigg)
\end{split}
\end{equation}
Using similar lines, one can also calculate the solution for $a(t)$ with the boundary condition $a(t=0) = - r_0^2/(2\theta)$. 
The solution is not reported here because it is very lengthy. 

We can now derive the average value on noise:
\begin{equation}
\left< \bar{C}(t) \right> = \sqrt{\frac{\pi}{c(t)}}\exp\left(a(t)+\frac{b^2(t)}{4c(t)}\right).
\end{equation}
We then find that $\left< \bar{C}(t) \right>$ behaves as:
\begin{equation}
\left< \bar{C}(t) \right> =\exp(-K_{\infty}(\omega)t + k_0(t)),
\end{equation}
where $k_0(t)$ is a small and periodic time contribution, which is sublinear in time and thus negligible for long time scales. $K_{\infty}(\omega)$ is the permeance and is such that:
\begin{equation}
\left< \bar{C}(t) \right>\underset{t\rightarrow \infty}{ \sim} \exp(-K_{\infty}(\omega)t),
\end{equation}
with
\begin{equation}
\label{muZwanzig1}
K_{\infty}(\omega) = \lambda/2(S-1) +  kr_0^2\left(\frac{1}{S^2} + \frac{\epsilon^2}{2}\frac{\lambda^2 + \omega^2}{S^2\lambda^2+\omega^2}\right).
\end{equation}
By replacing $S = (1 + \frac{4k'\theta}{\lambda})^{1/2}$ we find exactly the result of Eq.~(\ref{Kinf}).


\subsection*{Appendix C : Numerical methods and solutions}
\label{sectionNumerics}

\subsubsection*{C - 1 - Numerical methods}

The Schmoluchowski equations are solved with a finite difference scheme over 4 orders of magnitude of both $\omega$ and $\lambda$. Several methods are used to ensure global convergence:
\begin{itemize}
\item \textit{Change in space variable} We define $\tilde{C}(x,t) = \bar{C}(x - x_0(t),t)$ such that $\tilde{C}$ obeys a simpler Schmoluchowski equation (no variation of the drift coefficient in time) and solve for $\tilde{C}$ instead of $\bar{C}$. 
\item \textit{Logarithmic scale} We define and solve for ${\hat{C}} = \log\tilde{C}$. This yields a non linear equation, but the advantage is that high precision is gained - the initial condition is indeed a gaussian, and behaves much more nicely (on a smaller number of orders of magnitude) in gaussian scale. 
\item \textit{Partial Cranck-Nicholson} We perform a Cranck Nicholson scheme on the linear part, and explicit propagation on the non-linear part. Auto adaptative time scale is used to check for convergence in time. 
\item The initial time step is chosen via a burning algorithm that allows to adapt for any kind of parameters in the $(\omega,\lambda)$ parameter space. 
\item It was found that around a discretization of 1000 space steps usually gave reasonably convergent results. This number had to be adapted for different values in the parameter space anyway to ensure optimal convergence. 
\item $K_{\infty}$ was computed as an average over several periods (usually 10) of the relaxation rate, after an initial reasonably long transient phase. A very small amount of configurations (less than 10 over 100 points), with high $\lambda$ and small $\omega$, would relax to numerically untractable small concentrations before a single oscillation period expired. Averaging over several periods was thus impossible. The data obtained for these very few very small frequencies was equated with the values obtained for higher frequencies at the same $\lambda$, for plotting purposes. At these high $\lambda$, $K_{\infty}$ is not expected to depend on $\omega$. 
\end{itemize}

\subsubsection*{C - 2 - Systematic analysis in the absence of forcing}

\paragraph*{Limiting behaviors}

In the case of very high damping $\lambda \gg 1$, $x$ has almost the equilibrium distribution at all times, and thus the transition ability can be approximated by:
\begin{equation}
K_{\infty} \underset{\lambda \gg 1}{=} \int K(x) \rho_{eq}(x) dx
\label{highDamp}
\end{equation}
In the case of very low damping, we expect the following scaling discussed in the main paper:
\begin{equation}
K_{\infty} \underset{\lambda \ll 1}{=} K( \sqrt{\lambda/K_{\infty}} )
\label{LowDamp}
\end{equation}
Note that when the leakage law is a power law of the type $K(x) = |x|^n$, with $n$ some integer, then one easily finds 
\begin{equation}
K_{\infty} \sim \lambda^{\frac{n}{2+n}}
\end{equation} 
Equivalently, since the mobility $\lambda$ depends inversly on the viscosity $\eta$ of the fluid, $K_{\infty} \sim \eta^{-\frac{n}{2+n}}$.
\vskip0.2cm
In the following we check these scaling laws for different rate processes. The results are summarized within the following paragraphs.

\paragraph*{Quadratic rate process}

Correspondence between simulations of the quadratic rate process and its exact solution was verified as a benchmarl. 
We do not report this checking procedure here because it adds nothing to the discussion.

\paragraph*{Linear rate process}

\begin{figure}[h!]
\center
\includegraphics[width = \columnwidth]{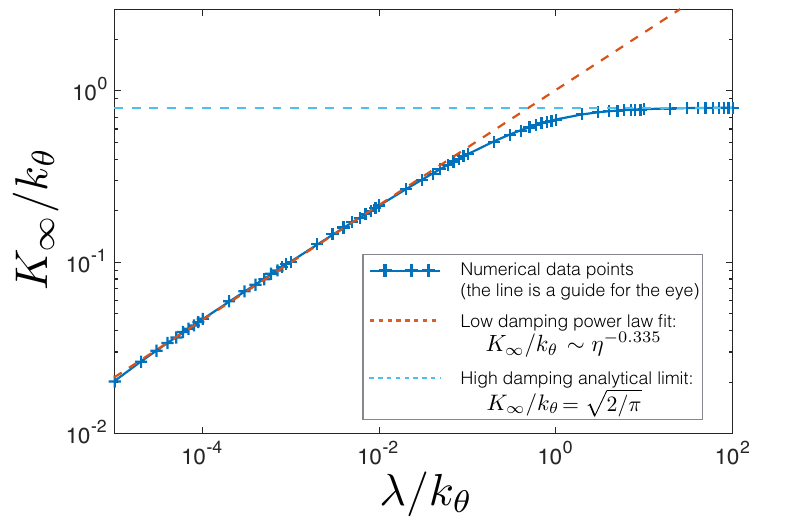}
\caption{\textbf{Linear rate process permeance in the absence of forcing (nanodoor)} Simulation results for the permeance $K_{\infty}$ as a function of the thermal damping $\lambda$ and comparison to analytical estimates. In this analysis $k_{\theta} = k \sqrt{\theta}$. $\eta$ is the viscosity of the fluid and is proportional to $1/\lambda$. }
\label{figS1}
\end{figure}
In the following paragraph we consider the leakage law associated typically with the nanodoor of Fig.~1a of the main text, $K(x) = k|x|$.
In Fig.~\ref{figS1} we show the permeance of the linear rate process as computed numerically. It verifies well the predicted low damping scaling law $K_{\infty} \sim \lambda^{1/3}$. 
The high damping limit is computed thanks to Eq.~(\ref{highDamp}) and is also in very good agreement with numerical calculations.

Note that in order to probe the previous scaling argument, $K_{\infty} \sim \lambda^{\frac{n}{2+n}}$, we also probed numerically other exponents in the leakage laws. For example 
for a cubic leakage law ($n = 3$), the numerical resolution yields an exponent of $0.58$, to compare with  the analytic estimate of  $3/(2+3)=0.6$. 

\paragraph*{Charge regulated rate process}

\begin{figure}[h!]
\center
\includegraphics[width=\columnwidth]{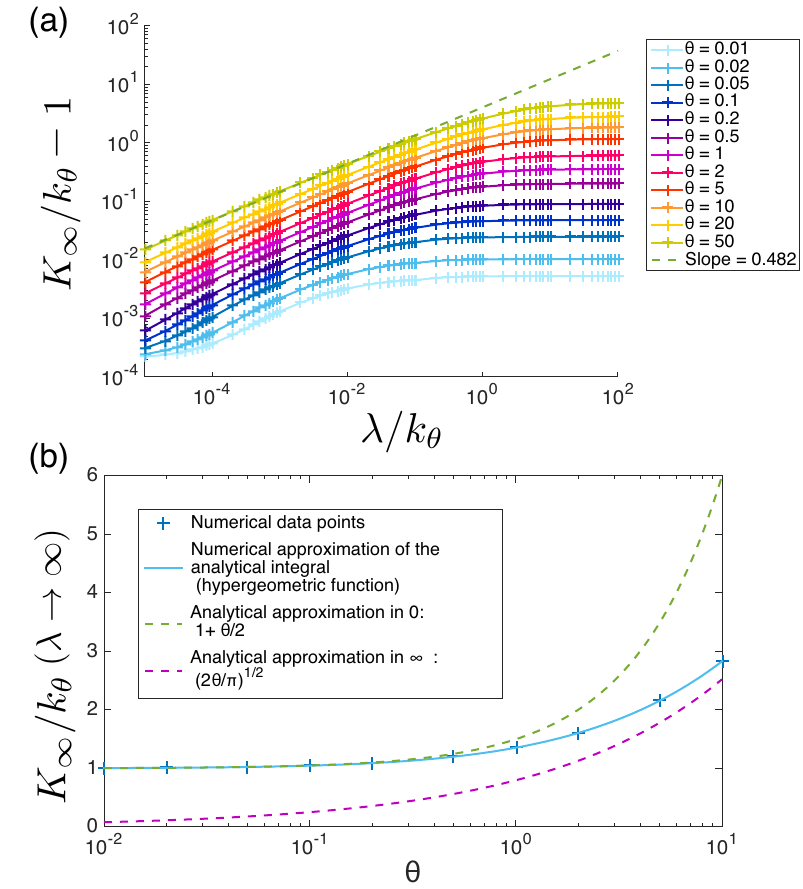}
\caption{\textbf{Charge regulated rate process permeance in the absence of forcing.} \textbf{(a)} Simulation results for the permeance $K_{\infty}$ as a function of the thermal damping $\lambda$ and comparison to analytical estimates. In this analysis $k_{\theta} = k''$. $\theta$ is kept in the derivation and varied. The dashed green line verifies $K_{\infty}/k_{\theta} - 1 \propto (\lambda/k_{\theta})^{\beta}$ with $\beta\simeq 0.482$. 
\textbf{(b)} Simulation results for the permeance $K_{\infty}$ as a function of $\theta$ at high thermal damping $\lambda$ and comparison to analytical estimates. }
\label{figS2}
\end{figure}

We finally consider the leakage law associated typically with the charged nanopore of Fig.~\ref{fig1}c of the main text, $K(\sigma) = k''\sqrt{1+\sigma^2}$.
In Fig.~\ref{figS2} we show the permeance of the charge regulated process. The process has more features because 
In the present case $\theta$ accounts for the fluctuations of the (dimensionless) surface charge $\sigma$. 
In Fig.~\ref{figS2}a we observe the permeance at different $\theta$ and find that for small $\lambda$ and large $\theta$, the system behaves as if it had an average leakage law behaving as $K(\sigma) \sim k'' + k''\sigma^2$, {\it i.e.} with an exponent $n=2$. This correspondance is not obvious a priori but allows to predict the scaling behavior for $K_{\infty}$. Indeed, one may write that typically the diffusion time scales like $1/(K_{\infty}-1)$, and one may then recover from Eq.~(\ref{LowDamp}) that $K_{\infty} - 1 \sim \lambda^{1/2}$. This scaling prediction us confirmed numerically, see Fig.~\ref{figS2}a. In the regime of high damping, see Fig.~\ref{figS2}b, we find that the system is well described by the analytical expression Eq.~(\ref{highDamp}) for any $\theta$.

\subsubsection*{C - 3 - Systematic analysis with forcing}

We now perform simulations with an external forcing at frequency $\omega$ and check that we get for the different systems (nanodoor and charged pore in Figs.~\ref{figS3}-\ref{figS4}) the same "universal features" as for the case of the circular pore. As for the circual nanoporen we indeed find 3 regimes: a forcing dominated regime, a noise dominated regime, and a global average regime, as described in the main text. 

We also want to check some analytic scaling laws on the rate process theory with forcing at frequency $\omega$. We have first considered the predicted scaling of the critical frequency with $\lambda$ (see main text):
\begin{equation}
\omega_c(\lambda) \sim \sqrt{\lambda}
\end{equation}
 We performed a fit of each simulation (at constant $\lambda$) of $K_{\infty}$ with the shape of a high pass filter (with a plateau at low frequencies) similar to the function $H$. We accordingly extract for each $\lambda$ the value of $\omega_c$ and then find the scaling law between $\lambda$ and $\omega_c$. Overall we have verified that for all systems the threshold frequency$\omega_c$ does obey this scaling low over a range of $\lambda \sim 10^{-3} \rightarrow 10^{-1}$. 

Furthermore, we can check that the plateau value for the permeance at high $\lambda$ matches the expected prediction assuming that $x$ reaches its equilibrium distribution (see Figs.~\ref{figS3}b). Since the equilibrium distribution of $x$ depends on $t$ (it is periodic over a period $T = \frac{2\pi}{\omega}$), we should also average over a period. This writes:
\begin{equation}
K_{\infty} \underset{\lambda \gg 1}{=} \left\langle \int K(x) \rho_{eq}(x-x_0(t)) dx \right\rangle_T
\label{HighLambda}
\end{equation}

\paragraph*{Quadratic rate process}

As above, the correspondence between simulations of the quadratic rate process at various frequencies and its exact solution was verified before moving on to cases not solvable analytically. This test procedure is not shown here because it does not add  to the discussion.

\paragraph*{Linear rate process}
In this paragraph we consider the leakage law associated typically with the nanodoor of Fig.~\ref{fig1}a of the main text, $K(x) = k|x|$.

In Fig.~\ref{figS3}a we show the permeance of the nanodoor system over 4 ranges of frequencies and damping. We find the three regimes of permeance (noise dominated regime in blue, forcing dominated regime in orange, and global average in yellow). The fitting procedure, described in Fig.~\ref{figS3}c, allows to find $\omega_c$ for each $\lambda$. In Fig.~\ref{figS3}d we plot $\omega_c$ as a function of $\sqrt{\lambda}$ and find a perfect agreement, that confirms the analytical prediction that $\omega_c \propto \sqrt{\lambda}$. The high plateau value (in yellow in Fig.~\ref{figS3}a) for various forcing amplitudes is shown in Fig.~\ref{figS3}b and agrees well with the prediction of Eq.~(\ref{HighLambda}).

\begin{figure}[h]
\center
\includegraphics[width=\columnwidth]{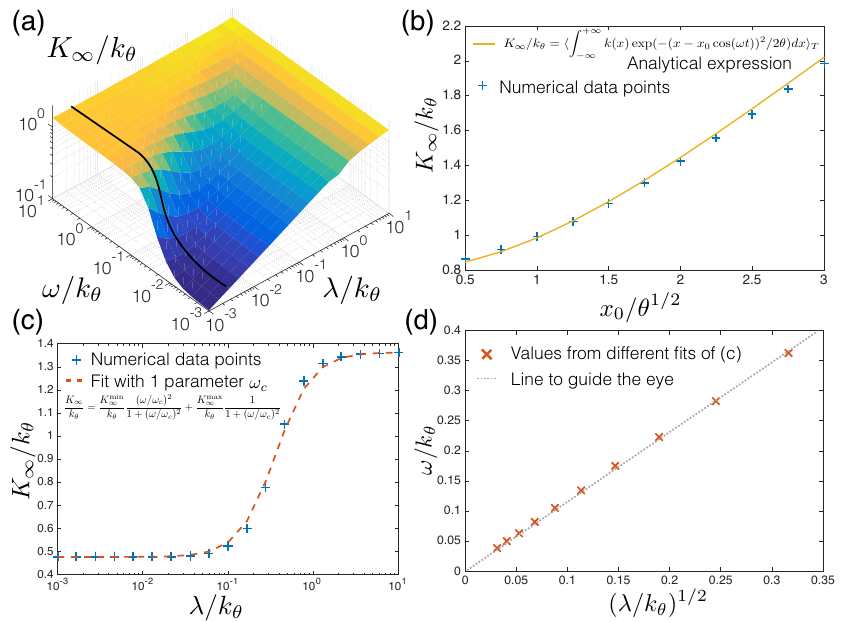}
\caption{\textbf{Linear rate process permeance} \textbf{(a)} Simulation results for the permeance $K_{\infty}$ as a function of the thermal damping $\lambda$ and the forcing frequency $\omega$ and comparison to analytical estimates. In this analysis $k_{\theta} = k \sqrt{\theta}$. The simulation parameter for oscillations around the origin is $x_0/\sqrt{\theta} = 2$. The black line indicates the fitting procedure at a given $\lambda$ to find $\omega_c$. \textbf{(b)} High plateau value for $\lambda/k_{\theta} = 10$ and $\omega/k_{\theta}= 0.1$ for different values of the ratio $x_0/\sqrt{\theta}$ where $x_0$ is the amplitude of the forcing $x_0(t) = x_0 \cos \omega t$.  \textbf{(c)} Example of a filter fit to determine $\omega_c$ from the permeance as a function of $\omega$ at a fixed $\lambda = 0.1 k_{\theta}$. The type of filter used is given in the legend and the only fitting parameter is $\omega_c$.  \textbf{(d)} Plot of $\omega_c$ as determined according to {\bf (c)} as a function of $\sqrt{\lambda}$. }
\label{figS3}
\end{figure}

\paragraph*{Charge regulated rate process}

Finally  we consider the leakage law associated typically with the charged nanopore of Fig.~\ref{fig1}c of the main text, $K(\sigma) = k''\sqrt{1+\sigma^2}$. 
In Fig.~\ref{figS4} we show the permeance of the charged pore over 4 orders of magnitude in frequencies and damping. We find the three regimes of permeance (noise dominated regime in blue, forcing dominated regime in orange, and global average in yellow). The fitting procedure yields a typical dependence $\omega_c(\lambda) \sim \lambda^{0.46}$, see Fig.~\ref{figS4}b very close to the analytical exponent ($0.5$). 

\begin{figure}[h]
\center
\includegraphics[width=\columnwidth]{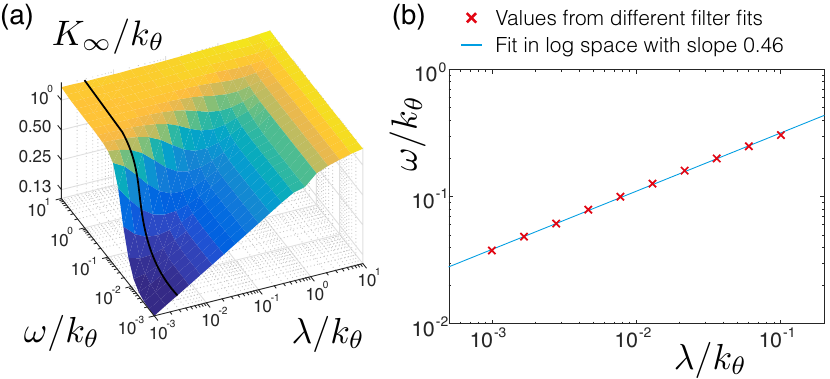}
\caption{\textbf{Charge regulated rate process permeance}  \textbf{(a)}~Simulation results for the permeance $K_{\infty}$ as a function of the thermal damping $\lambda$ and the forcing frequency $\omega$. In this analysis $k_{\theta} = k'' $. The simulation parameter for oscillations around the origin is $\sigma_0/\sqrt{\theta} = 2$, and $\theta = 1$. The black line indicates the fitting procedure at a given $\lambda$ to find $\omega_c$.  \textbf{(b)}~Plot of $\omega_c$ as determined from filter fits as a function of $\lambda$, and power law fit.}
\label{figS4}
\end{figure}

%

\end{document}